\documentclass[11pt]{article}
\usepackage{graphicx}
\usepackage{amssymb, amsmath}
\usepackage{amsfonts}
\usepackage{epstopdf}
\title{Generalizations of Quantum Mechanics}
\author{Philip Pearle\\ 
Hamilton College\\
 Clinton, NY 13323, USA\\
e-mail: ppearle@hamilton.edu
 \and 
 Antony Valentini \\
 Perimeter Institute for Theoretical Physics\\
31 Caroline Street North, Waterloo, Ontario N2L 2Y5, Canada\\
e-mail: avalentini@perimeterinstitute.ca}
\begin{document}
\date{}
\maketitle
\abstract{We review realistic models that reproduce quantum theory in some limit and
yield potentially new physics outside that limit. In particular, we consider
deterministic hidden-variables theories (such as the pilot-wave model) and
their extension to ``quantum nonequilibrium," and we consider the continuous 
spontaneous localization model of wave
function collapse. Other models are briefly discussed.}

\

\textbf{CONTENTS}

\ 

\textbf{1 Introduction}

\textbf{2 Hidden Variables and Quantum Nonequilibrium}

-- 2.1 Pilot-Wave Theory

-- 2.2 \textit{H}-Theorem: Relaxation to Equilibrium

-- 2.3 Nonlocal Signaling

-- 2.4 Subquantum Measurement

-- 2.5 Subquantum Information and Computation

-- 2.6 Extension to All Deterministic Hidden-Variables Theories

\textbf{3 Continuous Spontaneous Localization Model (CSL)}

-- 3.1 Requirements for Stochastic Collapse Dynamics

-- 3.2 CSL in Essence

-- 3.3 CSL

-- 3.4 Consequences of CSL

-- 3.5 Further Remarks

-- 3.6 Spontaneous Localization Model (SL)

\textbf{4 Other Models}

\bigskip

To be published in: \textit{Encyclopaedia of Mathematical Physics}, eds.
J.-P. Fran\c{c}oise, G. Naber and T. S. Tsun (Elsevier, 2006).

\section{Introduction}

	 According to the so-called ``Copenhagen Interpretation,"  standard quantum theory  is limited to describing experimental situations.  It is at once remarkably successful in its predictions, and remarkably ill-defined in its conceptual structure: what is an experiment?  what physical objects do or do not require quantization? how are the states realized in nature to be characterized?  how and when is the wave-function ``collapse postulate"  to be invoked? Because of its success, one may suspect that quantum theory can be promoted from a theory of measurement to a theory of reality.   But, that requires there to be an unambiguous specification (S) of the possible real states of Nature and their probabilities of being realized.  

	There are several approaches that attempt to achieve S.  The more conservative approaches (consistent histories, environmental decoherence, many worlds) do not  produce any predictions that differ from the standard ones  because they do not tamper with the usual basic mathematical formalism.    Rather, they utilize structures compatible with standard quantum theory to elucidate S.  These approaches, which will not be discussed in this article, have  arguably been less successful so far at achieving S than approaches that introduce significant alterations to quantum theory.  

	This article will largely deal with the two most well-developed realistic models that reproduce quantum theory in some limit and yield potentially new and testable physics outside that limit.  
First, the pilot-wave model, which will be discussed in the broader context of ``hidden-variables theories."   Second, the continuous spontaneous localization (CSL) model, which describes wave function collapse as a physical process. Other related models will also be discussed briefly. 

	Due to bibliographic space limitations, this article contains a number of uncited references, of the form ``(author) in (year)." Those in the next section can be found in Valentini (2002b, 2004a,b) or at www.arxiv.org.  Those in subsequent sections can be found in Adler (2004), Bassi and
Ghirardi (2003), Pearle (1999) (or in subsequent papers by
these authors, or directly, at www.arxiv.org), and in Wallstrom (1994).
	
\section{Hidden Variables and Quantum Nonequilibrium}

A deterministic hidden-variables theory defines a mapping $\omega =\omega
(M,\lambda )$ from initial hidden parameters $\lambda $ (defined, e.g., at
the time of preparation of a quantum state) to final outcomes $\omega $ of
quantum measurements. The mapping depends on macroscopic experimental
settings $M$, and fixes the outcome for each run of the experiment. Bell's
theorem of 1964 shows that, for entangled quantum states of widely separated
systems, the mapping must be nonlocal: some outcomes for (at least) one
system must depend on the setting for another distant system.

In a viable theory, the statistics of quantum measurement outcomes -- over
an ensemble of experimental trials with fixed settings $M$ -- will agree
with quantum theory for some special distribution $\rho _{\mathrm{QT}%
}(\lambda )$ of hidden variables. For example, expectation values will
coincide with the predictions of the Born rule%
\begin{equation*}
\left\langle \omega \right\rangle _{\mathrm{QT}}\equiv \int d\lambda \ \rho
_{\mathrm{QT}}(\lambda )\omega (M,\lambda )=\mathrm{Tr}(\hat{\rho}\hat{\Omega%
})
\end{equation*}%
for an appropriate density operator $\hat{\rho}$ and Hermitian observable $%
\hat{\Omega}$. (As is customary in this context, $\int d\lambda $ is to be
understood as a generalised sum.) However, given the mapping $\omega =\omega
(M,\lambda )$ for individual trials, one may, in principle, consider
nonstandard distributions $\rho (\lambda )\neq \rho _{\mathrm{QT}}(\lambda )$
that yield statistics outside the domain of ordinary quantum theory
(Valentini 1991, 2002a). We may say that such distributions correspond to a
state of quantum nonequilibrium.

Quantum nonequilibrium is characterised by the breakdown of a number of
basic quantum constraints. In particular, nonlocal signals appear at the
statistical level. We shall first illustrate this for the hidden-variables
model of de Broglie and Bohm. Then we shall generalize the discussion to all
(deterministic) hidden-variables theories.

At present there is no experimental evidence for quantum nonequilibrium in
nature. However, from a hidden-variables perspective, it is natural to
explore the theoretical properties of nonequilibrium distributions, and to
search experimentally for the statistical anomalies associated with them.

From this point of view, quantum theory is a special case of a wider
physics, much as thermal physics is a special case of a wider
(nonequilibrium) physics. (The special distribution $\rho _{\mathrm{QT}%
}(\lambda )$ is analogous to, say, Maxwell's distribution of molecular
speeds.) Quantum physics may be compared with the physics of global thermal
equilibrium, which is characterised by constraints -- such as the
impossibility of converting heat into work (in the absence of temperature
differences) -- that are not fundamental but contingent on the state.
Similarly, quantum constraints such as statistical locality (the
impossibility of converting entanglement into a practical signal) are seen
as contingencies of $\rho _{\mathrm{QT}}(\lambda )$.

\subsection{Pilot-Wave Theory}

The de Broglie-Bohm \textquotedblleft pilot-wave theory\textquotedblright\
-- as it was originally called by de Broglie, who first presented it at the
Fifth Solvay Congress in 1927 -- is the classic example of a deterministic
hidden-variables theory of broad scope (Bohm 1952; Bell 1987; Holland 1993).
We shall use it to illustrate the above ideas. Later, the discussion will be
generalised to arbitrary theories.

In pilot-wave dynamics, an individual closed system with
(configuration-space) wave function $\Psi (X,t)$ satisfying the Schr\"{o}%
dinger equation%
\begin{equation}
i\hbar \frac{\partial \Psi }{\partial t}=\hat{H}\Psi   \label{Sch0}
\end{equation}%
has an actual configuration $X(t)$ with velocity%
\begin{equation}
\dot{X}(t)=\frac{J(X,t)}{\left\vert \Psi (X,t)\right\vert ^{2}}  \label{deB0}
\end{equation}%
where $J=J\left[ \Psi \right] =J(X,t)$ satisfies the continuity equation%
\begin{equation}
\frac{\partial \left\vert \Psi \right\vert ^{2}}{\partial t}+\nabla \cdot J=0
\label{Cont0}
\end{equation}%
(which follows from (\ref{Sch0})). In quantum theory, $J$ is the
\textquotedblleft probability current\textquotedblright . In pilot-wave
theory, $\Psi $ is an objective physical field (on configuration space)
guiding the motion of an individual system.

Here, the objective state (or ontology) for a closed system is given by $%
\Psi $ and $X$. A probability distribution for $X$ -- discussed below --
completes an unambiguous specification S (as mentioned in the introduction).

Pilot-wave dynamics may be applied to any quantum system with a locally
conserved current in configuration space. Thus, $X$ may represent a
many-body system, or the configuration of a continuous field, or perhaps
some other entity.

For example, at low energies, for a system of $N$ particles with positions $%
\mathbf{x}_{i}(t)$ and masses $m_{i}$ ($i=1,2,....,N)$, with an external
potential $V$, (\ref{Sch0}) (with $X\equiv(\mathbf{x}_{1},\mathbf{x}%
_{2},....,\mathbf{x}_{N})$) reads%
\begin{equation}
i\hbar\frac{\partial\Psi}{\partial t}=\sum_{i=1}^{N}-\frac{\hbar^{2}}{2m_{i}}%
\nabla_{i}^{2}\Psi+V\Psi  \label{Sch1}
\end{equation}
while (\ref{deB0}) has components%
\begin{equation}
\frac{d\mathbf{x}_{i}}{dt}=\frac{\hbar}{m_{i}}\hbox{Im}\left( \frac{%
\mathbf{\nabla}_{i}\Psi}{\Psi}\right) =\frac{\mathbf{\nabla}_{i}S}{m_{i}}
\label{deB1}
\end{equation}
(where $\Psi=\left\vert \Psi\right\vert e^{(i/\hbar)S}$).

In general, (\ref{Sch0}) and (\ref{deB0}) determine $X(t)$ for an individual
system, given the initial condtions $X(0)$, $\Psi(X,0)$ at $t=0$. For an
arbitrary initial distribution $P(X,0)$, over an ensemble with the same wave
function $\Psi(X,0)$, the evolution $P(X,t)$ of the distribution is given by
the continuity equation%
\begin{equation}
\frac{\partial P}{\partial t}+\nabla\cdot(P\dot{X})=0  \label{Cont1}
\end{equation}

The outcome of an experiment is determined by $X(0)$, $\Psi(X,0)$, which may
be identified with $\lambda$. For an ensemble with the same $\Psi(X,0)$, we
have $\lambda=$ $X(0)$.

\subsubsection{Quantum Equilibrium}

From (\ref{Cont0}) and (\ref{Cont1}), if we assume $P(X,0)=\left\vert \Psi
(X,0)\right\vert ^{2}$ at $t=0$, we obtain $P(X,t)=\left\vert \Psi
(X,t)\right\vert ^{2}$ -- the Born-rule distribution of configurations -- at
all times $t$.

Quantum measurements are, like any other process, described and explained in
terms of evolving configurations. For measurement devices whose pointer
readings reduce to configurations, the distribution of outcomes of quantum
measurements will match the statistical predictions of quantum theory (Bohm
1952; Bell 1987; D\"{u}rr, Goldstein and Zangh\`{\i} 2003). Thus quantum
theory emerges phenomenologically for a \textquotedblleft quantum
equilibrium\textquotedblright\ ensemble with distribution $P(X,t)=\left\vert
\Psi (X,t)\right\vert ^{2}$ (or $\rho (\lambda )=\rho _{\mathrm{QT}}(\lambda
)$).

\subsubsection{Quantum Nonequilibrium}

In principle, as we saw for general hidden-variables theories, we may
consider a nonequilibrium distribution $P(X,0)\neq \left\vert \Psi
(X,0)\right\vert ^{2}$ of initial configurations while retaining the same
deterministic dynamics (\ref{Sch0}), (\ref{deB0}) for individual systems
(Valentini 1991). The time evolution of $P(X,t)$ will be determined by (\ref%
{Cont1}).

As we shall see, in appropriate circumstances (with a sufficiently
complicated velocity field $\dot{X}$), (\ref{Cont1}) generates relaxation $%
P\rightarrow \left\vert \Psi \right\vert ^{2}$ on a coarse-grained level,
much as the analogous classical evolution on phase space generates thermal
relaxation. But for as long as the ensemble is in nonequilibrium, the
statistics of outcomes of quantum measurements will disagree with quantum
theory.

Quantum nonequilibrium may have existed in the very early universe, with
relaxation to equilibrium occurring soon after the big bang. Thus, a
hidden-variables analogue of the classical thermodynamic \textquotedblleft
heat death of the universe\textquotedblright\ may have actually taken place
(Valentini 1991). Even so, relic cosmological particles that decoupled
sufficiently early could still be in nonequilibrium today, as suggested by
Valentini in 1996 and 2001. It has also been speculated that nonequilibrium
could be generated in systems entangled with degrees of freedom behind a
black-hole event horizon (Valentini 2004a).

Experimental searches for nonequilibrium have been proposed. Nonequilibrium
could be detected by the statistical analysis of random samples of particles
taken from a parent population of (for example) relics from the early
universe. Once the parent distribution is known, the rest of the population
could be used as a resource, to perform tasks that are currently impossible
(Valentini 2002b).

\subsection{\textit{H}-Theorem: Relaxation to Equilibrium}

Before discussing the potential uses of nonequilibrium, we should first
explain why all systems probed so far have been found in the equilibrium
state $P=\left\vert \Psi \right\vert ^{2}$. This distribution may be
accounted for along the lines of classical statistical mechanics, noting
that all currently-accessible systems have had a long and violent
astrophysical history.

Dividing configuration space into small cells, and introducing
coarse-grained quantities $\bar{P}$, $\overline{\left\vert \Psi \right\vert
^{2}}$, a general argument for relaxation $\bar{P}\rightarrow \overline{%
\left\vert \Psi \right\vert ^{2}}$ is based on an analog of the classical
coarse-graining \textit{H}-theorem. The coarse-grained \textit{H}-function%
\begin{equation}
\bar{H}=\int dX\;\bar{P}\ln (\bar{P}/\overline{\left\vert \Psi \right\vert
^{2}})  \label{Hbar}
\end{equation}%
(minus the relative entropy of $\bar{P}$ with respect to $\overline{%
\left\vert \Psi \right\vert ^{2}}$) obeys the \textit{H}-theorem (Valentini
1991)%
\begin{equation*}
\bar{H}(t)\leq \bar{H}(0)
\end{equation*}%
(assuming no initial fine-grained microstructure in $P$ and $\left\vert \Psi
\right\vert ^{2}$). Here, $\bar{H}\geq 0$ for all $\bar{P}$, $\overline{%
\left\vert \Psi \right\vert ^{2}}$ and $\bar{H}=0$ if and only if $\bar{P}=%
\overline{\left\vert \Psi \right\vert ^{2}}$ everywhere.

The \textit{H}-theorem expresses the fact that $P$ and $\left\vert \Psi
\right\vert ^{2}$ behave like two \textquotedblleft
fluids\textquotedblright\ that are \textquotedblleft
stirred\textquotedblright\ by the same velocity field $\dot{X}$, so that $P$
and $\left\vert \Psi \right\vert ^{2}$ tend to become indistinguishable on a
coarse-grained level. Like its classical analog, the theorem provides a
general understanding of how equilibrium is approached, while not proving
that equilibrium is actually reached. (And of course, for some simple
systems -- such as a particle in the ground state of a box, for which the
velocity field $\nabla S/m$ vanishes -- there is no relaxation at all.) A
strict decrease of $\bar{H}(t)$ immediately after $t=0$ is guaranteed if $%
\dot{X}_{0}\cdot \nabla (P_{0}/\left\vert \Psi _{0}\right\vert ^{2})$ has
non-zero spatial variance over a coarse-graining cell, as shown by Valentini
in 1992 and 2001.

A relaxation timescale $\tau $ may be defined by $1/\tau ^{2}\equiv -\left(
d^{2}\bar{H}/dt^{2}\right) _{0}/\bar{H}_{0}$. For a single particle with
quantum energy spread $\Delta E$, a crude estimate given by Valentini in
2001 yields $\tau \sim (1/\varepsilon )\hbar ^{2}/m^{1/2}(\Delta E)^{3/2}$,
where $\varepsilon $ is the coarse-graining length. For wave functions that
are superpositions of many energy eigenfunctions, the velocity field
(generally) varies rapidly, and detailed numerical simulations (in two
dimensions) show that relaxation occurs with an approximately exponential
decay $\bar{H}(t)\approx \bar{H}_{0}e^{-t/t_{c}}$, with a time constant $%
t_{c}$ of order $\tau $ (Valentini and Westman 2005).

Equilibrium is then to be expected for particles emerging from the violence
of the big bang. The possibility is still open that relics from very early
times may not have reached equilibrium before decoupling.

\subsection{Nonlocal Signaling}

We now show how nonequilibrium, if it were ever discovered, could be used
for nonlocal signaling.

Pilot-wave dynamics is nonlocal. For a pair of particles $A$, $B$ with
entangled wave function $\Psi(\mathbf{x}_{A},\mathbf{x}_{B},t)$, the
velocity $\mathbf{\dot{x}}_{A}(t)=\mathbf{\nabla}_{A}S(\mathbf{x}_{A},%
\mathbf{x}_{B},t)/m_{A}$ of $A$ depends instantaneously on $\mathbf{x}_{B}$,
and local operations at $B$ -- such as switching on a potential --
instantaneously affect the motion of $A$. For an ensemble $P(\mathbf{x}_{A},%
\mathbf{x}_{B},t)=|\Psi(\mathbf{x}_{A},\mathbf{x}_{B},t)|^{2}$, local
operations at $B$ have no statistical effect at $A$: the individual nonlocal
effects vanish upon averaging over an equilibrium ensemble.

Nonlocality is (generally) hidden by statistical noise only in quantum
equilibrium. If instead $P(\mathbf{x}_{A},\mathbf{x}_{B},0)\neq |\Psi (%
\mathbf{x}_{A},\mathbf{x}_{B},0)|^{2}$, a local change in the Hamiltonian at 
$B$ generally induces an instantaneous change in the marginal $p_{A}(\mathbf{%
x}_{A},t)\equiv \int d^{3}\mathbf{x}_{B}\ P(\mathbf{x}_{A},\mathbf{x}_{B},t)
$ at $A$. For example, in one dimension a sudden change $\hat{H}%
_{B}\rightarrow \hat{H}_{B}^{\prime }$ in the Hamiltonian at $B$ induces a
change $\Delta p_{A}\equiv p_{A}(x_{A},t)-p_{A}(x_{A},0)$ (for small $t$)
(Valentini 1991),%
\begin{equation}
\Delta p_{A}=-\frac{t^{2}}{4m}\frac{\partial }{\partial x_{A}}\left(
a(x_{A})\int dx_{B}\ b(x_{B})\frac{P(x_{A},x_{B},0)-|\Psi
(x_{A},x_{B},0)|^{2}}{|\Psi (x_{A},x_{B},0)|^{2}}\right)   \label{Superc}
\end{equation}%
(Here $m_{A}=m_{B}=m$, $a(x_{A})$ depends on $\Psi (x_{A},x_{B},0)$, while $%
b(x_{B})$ also depends on $\hat{H}_{B}^{\prime }$ and vanishes if $\hat{H}%
_{B}^{\prime }=\hat{H}_{B}$.) The signal is generally nonzero if $P_{0}\neq
|\Psi _{0}|^{2}$.

Nonlocal signals do not lead to causal paradoxes if, at the hidden-variable
level, there is a preferred foliation of spacetime with a time parameter
that defines a fundamental causal sequence. Such signals, if they were
observed, would define an absolute simultaneity as discussed by Valentini in
1992 and 2005. Note that in pilot-wave field theory, Lorentz invariance
emerges as a phenomenological symmetry of the equilibrium state, conditional
on the structure of the field-theoretical Hamiltonian (as discussed by Bohm
and Hiley in 1984, Bohm, Hiley and Kaloyerou in 1987, and Valentini in 1992
and 1996).

\subsection{Subquantum Measurement}

In principle, nonequilibrium particles could also be used to perform
\textquotedblleft subquantum measurements\textquotedblright\ on ordinary,
equilibrium systems. We illustrate this with an exactly solvable
one-dimensional model (Valentini 2002b).

Consider an apparatus \textquotedblleft pointer\textquotedblright\
coordinate $y$, with known wave function $g_{0}(y)$ and known (ensemble)
distribution $\pi _{0}(y)\neq \left\vert g_{0}(y)\right\vert ^{2}$, where $%
\pi _{0}(y)$ has been deduced by statistical analysis of random samples from
a parent population with known wave function $g_{0}(y)$. (We assume that
relaxation may be neglected: for example, if $g_{0}$ is a box ground state, $%
\dot{y}=0$ and $\pi _{0}(y)$ is static.) Consider also a \textquotedblleft
system\textquotedblright\ coordinate $x$ with known wave function $\psi
_{0}(x)$ and known distribution $\rho _{0}(x)=\left\vert \psi
_{0}(x)\right\vert ^{2}$. If $\pi _{0}(y)$ is arbitrarily narrow, $x_{0}$
can be measured without disturbing $\psi _{0}(x)$, to arbitrary accuracy
(violating the uncertainty principle).

To do this, at $t=0$ we switch on an interaction Hamiltonian $\hat{H}=a%
\hat
{x}\hat{p}_{y}$, where $a$ is a constant and $p_{y}$ is canonically
conjugate to $y$. For relatively large $a$, we may neglect the Hamiltonians
of $x$ and $y$. For $\Psi=\Psi(x,y,t)$, we then have $\partial\Psi/\partial
t=-ax\partial\Psi/\partial y$. For $\left\vert \Psi\right\vert ^{2}$ we have
the continuity equation $\partial\left\vert \Psi\right\vert ^{2}/\partial
t=-ax\partial\left\vert \Psi\right\vert ^{2}/\partial y$, which implies the
hidden-variable velocity fields $\dot{x}=0,\;\dot{y}=ax$ and trajectories $%
x(t)=x_{0},\;y(t)=y_{0}+ax_{0}t$.

The initial product $\Psi _{0}(x,y)=\psi _{0}(x)g_{0}(y)$ evolves into $\Psi
(x,y,t)=\psi _{0}(x)g_{0}(y-axt)$. For $at\rightarrow 0$ (with $a$ large but
fixed), $\Psi (x,y,t)\rightarrow \psi _{0}(x)g_{0}(y)$ and $\psi _{0}(x)$ is
undisturbed: for small $at$, a standard quantum pointer with the coordinate $%
y$ would yield negligible information about $x_{0}$. Yet, for arbitrarily
small $at$, the hidden-variable pointer coordinate $y(t)=y_{0}+ax_{0}t$ does
contain complete information about $x_{0}$ (and $x(t)=x_{0}$). This
\textquotedblleft subquantum\textquotedblright\ information will be visible
to us if $\pi _{0}(y)$ is sufficiently narrow.

For, over an ensemble of similar experiments, with initial joint
distribution $P_{0}(x,y)=\left\vert \psi _{0}(x)\right\vert ^{2}\pi _{0}(y)$
(equilibrium for $x$ and nonequilibrium for $y$), the continuity equation $%
\partial P/\partial t=-ax\partial P/\partial y$ implies that $%
P(x,y,t)=\left\vert \psi _{0}(x)\right\vert ^{2}\pi _{0}(y-axt)$. If $\pi
_{0}(y)$ is localised around $y=0$ ($\pi _{0}(y)=0$ for $\left\vert
y\right\vert >w/2$), then a standard (faithful) measurement of $y$ with
result $y_{\mathrm{meas}}$ will imply that $x$ lies in the interval $(y_{%
\mathrm{meas}}/at-w/2at,\;y_{\mathrm{meas}}/at+w/2at)$ (so that $%
P(x,y,t)\neq 0$). Taking the simultaneous limits $at\rightarrow 0$, $%
w\rightarrow 0$, with $w/at\rightarrow 0$, the midpoint $y_{\mathrm{meas}%
}/at\rightarrow x_{0}$ (since $y_{\mathrm{meas}}=y_{0}+ax_{0}t$ and $%
\left\vert y_{0}\right\vert \leq w/2$), while the error $w/2at\rightarrow 0$.

If $w$ is arbitrarily small, a sequence of such measurements will determine
the hidden trajectory $x(t)$ without disturbing $\psi(x,t)$, to arbitrary
accuracy.

\subsection{Subquantum Information and Computation}

From a hidden-variables perspective, immense physical resources are hidden
from us by equilibrium statistical noise. Quantum nonequilibrium would
probably be as useful technologically as thermal or chemical nonequilibrium.

\subsubsection{Distinguishing Nonorthogonal States}

In quantum theory, nonorthogonal states $\left\vert \psi _{1}\right\rangle $%
, $\left\vert \psi _{2}\right\rangle $ ($\langle \psi _{1}|\psi _{2}\rangle
\neq 0$) cannot be distinguished without disturbing them. This theorem
breaks down in quantum nonequilibrium (Valentini 2002b). For example, if $%
\left\vert \psi _{1}\right\rangle $, $\left\vert \psi _{2}\right\rangle $
are distinct states of a single spinless particle, then the associated de
Broglie-Bohm velocity fields will in general be different, even if $\langle
\psi _{1}|\psi _{2}\rangle \neq 0$, and so will the hidden-variable
trajectories. Subquantum measurement of the trajectories could then
distinguish the states $\left\vert \psi _{1}\right\rangle $, $\left\vert
\psi _{2}\right\rangle $.

\subsubsection{Breaking Quantum Cryptography}

The security of standard protocols for quantum key distribution depends on
the validity of the laws of quantum theory. These protocols would become
insecure given the availability of nonequilibrium systems (Valentini 2002b).

The protocols known as BB84 and B92 depend on the impossibility of
distinguishing nonorthogonal quantum states without disturbing them. An
eavesdropper in possession of nonequilibrium particles could distinguish the
nonorthogonal states being transmitted between two parties, and so read the
supposedly secret key. Further, if subquantum measurements allow an
eavesdropper to predict quantum measurement outcomes at each
\textquotedblleft wing\textquotedblright\ of a (bipartite) entangled state,
then the EPR (Einstein-Podolsky-Rosen) protocol also becomes insecure.

\subsubsection{Subquantum Computation}

It has been suggested that nonequilibrium physics would be computationally
more powerful than quantum theory, because of the ability to distinguish
nonorthogonal states (Valentini 2002b). However, this ability depends on the
(less-than-quantum) dispersion $w$ of the nonequilibrium ensemble. A
well-defined model of computational complexity requires that the resources
be quantified in some way. Here, a key question is how the required $w$
scales with the size of the computational task. So far, no rigorous results
are known.

\subsection{Extension to All Deterministic Hidden-Variables Theories}

Let us now discuss arbitrary (deterministic) theories.

\subsubsection{Nonlocal Signaling}

Consider a pair of two-state quantum systems $A$ and $B$, which are widely
separated and in the singlet state. Quantum measurements of observables 
$\hat{\sigma}_{A}\equiv \mathbf{m}_{A}\cdot{\hat{\boldsymbol\sigma}}_{A}$, $\hat{%
\sigma}_{B}\equiv \mathbf{m}_{B}\cdot {\hat{\boldsymbol\sigma}}_{B}$ (where $%
\mathbf{m}_{A}$, $\mathbf{m}_{B}$ are unit vectors in Bloch space and $%
{\hat{\boldsymbol\sigma}}_{A}$, ${\hat{\boldsymbol\sigma}}_{B}$ are Pauli spin
operators) yield outcomes $\sigma _{A}$, $\sigma _{B}=\pm 1$, in the ratio $%
1:1$ at each wing, with a correlation $\left\langle \hat{\sigma}_{A}\hat{%
\sigma}_{B}\right\rangle =-\mathbf{m}_{A}\cdot \mathbf{m}_{B}$. Bell's
theorem shows that for a hidden-variables theory to reproduce this
correlation -- upon averaging over an equilibrium ensemble with distribution 
$\rho _{\mathrm{QT}}(\lambda )$ -- it must take the nonlocal form%
\begin{equation}
\sigma _{A}=\sigma _{A}(\mathbf{m}_{A},\mathbf{m}_{B},\lambda
),\;\;\;\;\sigma _{B}=\sigma _{B}(\mathbf{m}_{A},\mathbf{m}_{B},\lambda )
\label{deteqns}
\end{equation}%
More precisely, to obtain $\left\langle \sigma _{A}\sigma _{B}\right\rangle
_{\mathrm{QT}}=-\mathbf{m}_{A}\cdot \mathbf{m}_{B}$ (where $\left\langle
\sigma _{A}\sigma _{B}\right\rangle _{\mathrm{QT}}\equiv \int d\lambda \
\rho _{\mathrm{QT}}(\lambda )\sigma _{A}\sigma _{B}$), at least one of $%
\sigma _{A}$, $\sigma _{B}$ must depend on the measurement setting at the
distant wing. Without loss of generality, we assume that $\sigma _{A}$
depends on $\mathbf{m}_{B}$.

For an arbitrary nonequilibrium ensemble with distribution $\rho (\lambda
)\neq \rho _{\mathrm{QT}}(\lambda )$, in general $\left\langle \sigma
_{A}\sigma _{B}\right\rangle \equiv \int d\lambda \ \rho (\lambda )\sigma
_{A}\sigma _{B}$ differs from $-\mathbf{m}_{A}\cdot \mathbf{m}_{B}$, and the
outcomes $\sigma _{A}$, $\sigma _{B}=\pm 1$ occur in a ratio different from $%
1:1$. Further, a change of setting $\mathbf{m}_{B}\rightarrow \mathbf{m}%
_{B}^{\prime }$ at $B$ will generally induce a change in the outcome
statistics at $A$, yielding a nonlocal signal at the statistical level. To
see this, note that, in a nonlocal theory, the \textquotedblleft transition
sets\textquotedblright 
\begin{align*}
T_{A}(-,+)& \equiv \left\{ \lambda |\sigma _{A}(\mathbf{m}_{A},\mathbf{m}%
_{B},\lambda )=-1,\;\sigma _{A}(\mathbf{m}_{A},\mathbf{m}_{B}^{\prime
},\lambda )=+1\right\}  \\
T_{A}(+,-)& \equiv \left\{ \lambda |\sigma _{A}(\mathbf{m}_{A},\mathbf{m}%
_{B},\lambda )=+1,\;\sigma _{A}(\mathbf{m}_{A},\mathbf{m}_{B}^{\prime
},\lambda )=-1\right\} 
\end{align*}%
cannot be empty for arbitrary settings. Yet, in quantum equilibrium, the
outcomes $\sigma _{A}=\pm 1$ occur in the ratio $1:1$ for all settings, so
the transition sets must have equal equilibrium measure, $\mu _{\mathrm{QT}%
}[T_{A}(-,+)]=\mu _{\mathrm{QT}}[T_{A}(+,-)]$ ($d\mu _{\mathrm{QT}}\equiv
\rho _{\mathrm{QT}}(\lambda )d\lambda $). That is, the fraction of the
equilibrium ensemble making the transition $\sigma _{A}=-1\rightarrow \sigma
_{A}=+1$ under $\mathbf{m}_{B}\rightarrow \mathbf{m}_{B}^{\prime }$ must
equal the fraction making the reverse transition $\sigma _{A}=+1\rightarrow
\sigma _{A}=-1$. (This \textquotedblleft detailed
balancing\textquotedblright\ is analogous to the principle of detailed
balance in statistical mechanics.) Since $T_{A}(-,+)$, $T_{A}(+,-)$ are
fixed by the deterministic mapping, they are independent of the ensemble
distribution $\rho (\lambda )$. Thus, for $\rho (\lambda )\neq \rho _{%
\mathrm{QT}}(\lambda )$, in general $\mu \lbrack T_{A}(-,+)]\neq \mu \lbrack
T_{A}(+,-)]$ ($d\mu \equiv \rho (\lambda )d\lambda $): the fraction of the
nonequilibrium ensemble making the transition $\sigma _{A}=-1\rightarrow
\sigma _{A}=+1$ will not in general balance the fraction making the reverse
transition. The outcome ratio at $A$ will then change under $\mathbf{m}%
_{B}\rightarrow \mathbf{m}_{B}^{\prime }$ and there will be an instantaneous
signal at the statistical level from $B$ to $A$ (Valentini 2002a).

Thus, in any deterministic hidden-variables theory, nonequilibrium
distributions $\rho (\lambda )\neq \rho _{\mathrm{QT}}(\lambda )$ generally
allow entanglement to be used for nonlocal signaling (just as, in ordinary
statistical physics, differences of temperature make it possible to convert
heat into work).

\subsubsection{Experimental Signature of Nonequilibrium}

Quantum expectations are additive, $\langle c_{1}\hat{\Omega}_{1}+c_{2}\hat{%
\Omega}_{2}\rangle =c_{1}\langle \hat{\Omega}_{1}\rangle +c_{2}\langle \hat{%
\Omega}_{2}\rangle $, even for noncommuting observables ($[\hat{\Omega}_{1},%
\hat{\Omega}_{2}]\neq 0$, with $c_{1}$, $c_{2}$ real). As emphasised by Bell
in 1966, this seemingly trivial consequence of the (linearity of the) Born
rule $\langle \hat{\Omega}\rangle =\mathrm{Tr}(\hat{\rho}\hat{\Omega})$ is
remarkable because it relates statistics from distinct, \textquotedblleft
incompatible\textquotedblright\ experiments. In nonequilibrium, such
additivity generically breaks down (Valentini 2004b).

Further, for a two-state system with observables $\mathbf{m}\cdot {\hat{\boldsymbol\sigma}}$, the \textquotedblleft dot-product\textquotedblright\
structure of the quantum expectation $\left\langle \mathbf{m}\cdot {\hat{\boldsymbol\sigma}}\right\rangle =\mathrm{Tr}\left( \hat{\rho}\mathbf{m}\cdot 
{\hat{\boldsymbol\sigma}}\right) =\mathbf{m}\cdot \mathbf{P}$ (for some Bloch
vector $\mathbf{P}$) is equivalent to expectation additivity (Valentini
2004b). Nonadditive expectations then provide a convenient signature of
nonequilibrium for any two-state system. For example, the sinusoidal
modulation of the quantum transmission probability for a single photon
through a polariser%
\begin{equation}
p_{\mathrm{QT}}^{+}(\Theta )=\frac{1}{2}\left( 1+\left\langle \mathbf{m}%
\cdot{\hat{\boldsymbol\sigma}}\right\rangle \right) =\frac{1}{2}\left( 1+P\cos
2\Theta \right)   \label{Photcos}
\end{equation}%
(where an angle $\theta $ on the Bloch sphere corresponds to a physical
angle $\Theta =\theta /2$) will generically break down in nonequilibrium.
Deviations from (\ref{Photcos}) would provide an unambiguous violation of
quantum theory (Valentini 2004b).

Such deviations were searched for by Papaliolios in 1967, using laboratory
photons and successive polarisation measurements over very short times, to
test a hidden-variables theory (distinct from pilot-wave theory) due to Bohm
and Bub (1966), in which quantum measurements generate nonequilibrium for
short times. Experimentally, successive measurements over timescales $\sim
10^{-13}$ $\mathrm{\sec }$ agreed with the (quantum) sinusoidal modulation $%
\cos ^{2}\Theta $ to $\lesssim 1\%$. Similar tests might be performed with
photons of a more exotic origin.

\section {Continuous Spontaneous Localization Model (CSL)}\label{CSL}

	The basic postulate of CSL is that the state vector $|\psi,t\rangle$ represents reality.  Since, for example, in describing a measurement, the usual Schr\"odinger evolution readily takes a real state into a non-real state, that is, into a superposition of real states (such as apparatus states describing different experimental outcomes), CSL requires a modification of Schr\"odinger's  evolution.  To the Hamiltonian is added a term which depends upon a classical randomly fluctuating field $w({\bf x},t)$ and a mass-density operator $\hat A({\bf x},t)$.  This  term acts to collapse a superposition of states, which differ in their spatial distribution of mass density, to one of these states.  The rate of collapse is very slow for a superposition involving a few particles, but very fast for a superposition of macroscopically different states.  Thus, very rapidly, what you see (in nature) is what you get (from the theory).   Each state vector evolving under each $w({\bf x},t)$ corresponds to a realizable state, and a rule is given for how to associate a probability with each.  In this way, an unambiguous specification S as mentioned in the introduction is achieved.  
	
\subsection{Requirements for Stochastic Collapse Dynamics}	
	 	
	Consider a normalized state vector $|\psi, t\rangle=\sum_{n}\alpha_{n}(t)|a_{n}\rangle$ ($\langle a_{n}|a_{n'}\rangle=\delta_{nn'}$) which undergoes a stochastic dynamical collapse process. This means that, starting from the initial superposition at $t=0$, for each run of the process, the squared amplitudes $x_{n}(t)\equiv|\alpha_{n}(t)|^{2}$ fluctuate until all but one vanish, that is,  
$x_{m}(\infty)=1$, ($x_{\neq m}(\infty)=0)$ with probability $x_{m}(0)$.

	This may be achieved simply, assuming negligible effect of the usual Schr\"odinger evolution, if the stochastic process enjoys the following properties (Pearle 1979):
\begin{subequations}
\begin{align}	
\sum_{n}x_{n}(t)&=1\label{E1a}\\
\overline{x_{n}(t)}&=x_{n}(0)\label{E1b}\\
\overline{x_{n}(\infty)x_{m}(\infty)}&= 0\quad\mbox{for}\quad  m\neq n, \label{E1c}
\end{align}
\end{subequations}	
\noindent where the overbar indicates the ensemble average at the indicated time.  The only way that a sum of products of non-negative terms can vanish is for at least one term in each product to vanish.  Thus, according to (\ref{E1c}),  for  each run, 
at least one of  each pair \{$x_{n}(\infty)$, $x_{m}(\infty)$\} ($n\neq m$) must vanish.  This means that at most one $x_{n}(\infty)$  might not vanish and, by (\ref{E1a}), applied at $t=\infty$, one $x_{n}(\infty)$   must not vanish and, in fact, must equal 1:  hence, each run produces collapse.  Now,  let the probability of the outcome \{$x_{n}(\infty)=1$, $x_{\neq n}(\infty)=0$\}  be denoted $P_{n}$.  Since $\overline{x_{n}(\infty)}=1\cdot P_{n}+\sum_{m\neq n}0\cdot P_{m} =P_{n}$ then, according to the Martingale property (\ref{E1b}), 	applied at $t=\infty$,
$P_{n}=x_{n}(0)$: hence the ensemble of runs produces the probability postulated by the usual ``collapse rule" of standard quantum theory.  

A (nonquantum) stochastic process which obeys  these equations is the gambler's ruin game.  Suppose one gambler initially possesses the fraction $x_{1}(0)$ of their joint wealth, and the other has 
the fraction $x_{2}(0)$.  They toss a coin: heads, a dollar goes from gambler 1 to gambler 2, tails the dollar goes the other way.  (\ref{E1a})	is satisfied since the sum of money in the game remains constant, (\ref{E1b}) holds because it is a fair game, and  (\ref{E1c}) holds because each game eventually ends.  Thus, gambler $i$ wins all the money  with probability  $x_{i}(0)$.

\subsection{CSL in Essence}

	Consider the (non-unitary) Schr\"odinger picture evolution equation 
\begin{equation}\label{E2}
|\psi,t\rangle_{w}={\cal T}e^{-\int_{0}^{t}dt'\{i\hat H+(4\lambda)^{-1}[w(t')-2\lambda \hat A]^{2}\}}|\psi,0\rangle, 
\end{equation}
\noindent where $\hat H$ is the usual Hamiltonian, $w(t')$ is an arbitrary function of white noise class, $\hat A$ is a Hermitian operator ($\hat A|a_{n}\rangle=a_{n}|a_{n}\rangle$), $\lambda$ is a collapse rate parameter, ${\cal T}$ is the time-ordering operator and  $\hbar=1$.  Associated with this,  the probability rule 
\begin{equation}\label{E3}
P_{t}(w)Dw\equiv _{w}\negthinspace\negthinspace\negthinspace\langle \psi,t |\psi,t \rangle_{w}\prod_{j=0}^{t/dt}dw(t_{j})/(2\pi \lambda /dt)^{1/2}
\end{equation}
\noindent is defined, which gives the probability that nature chooses a noise which lies in the range 
$\{w(t'), w(t')+dw(t')\}$ (for calculational purposes, time is discretized, with $t_{0}=0$).  

Equations (\ref{E2}) and  (\ref{E3}) contain the essential features of CSL,  and are 
all that is needed to discuss the simplest collapse behavior.  Set $\hat H=0$, so there is no competition between collapse and the usual Schr\"odinger evolution, and let the initial statevector be 
$|\psi, 0\rangle=\sum_{n}\alpha_{n}|a_{n}\rangle$. Equations (\ref{E2}) and (\ref{E3}) become
\begin{subequations}\label{E4}
\begin{eqnarray}	
|\psi,t\rangle_{w}&=&\sum_{n}\alpha_{n}|a_{n}\rangle e^{-(4\lambda)^{-1}\int_{0}^{t}dt'[w(t')-2\lambda a_{n}]^{2}}\label{E4a}\\
P_{t}(w)&=&\sum_{n}|\alpha_{n}|^{2}e^{-(2\lambda)^{-1}\int_{0}^{t}dt' [w(t')-2\lambda a_{n}]^{2}}\label{E4b}.
\end{eqnarray}
\end{subequations}
\noindent When the un-normalized state vector in (\ref{E4a}) is 
divided by $P_{t}^{1/2}(w)$ and so normalized, the squared amplitudes are 
\[
x_{n}(t)=|\alpha_{n}|^{2}e^{-(2\lambda)^{-1}\int_{0}^{t}dt' [w(t')-2\lambda a_{n}]^{2}}/P_{t}(w), 
\]
which are readily shown to satisfy (\ref{E1a}), (\ref{E1b}), and (\ref{E1c}) in the form $\overline{x_{n}^{1/2}(\infty)x_{m}^{1/2}(\infty)}=0$ ($m\neq n$) (which does not change the argument in section 3.1, but makes for an easier calculation).  Thus, (\ref{E4a}), (\ref{E4b}) describe collapse dynamics.  

To describe collapse to a joint eigenstate of a set of mutually commuting operators $\hat A^{r}$, replace  $(4\lambda)^{-1}[w(t')-2\lambda \hat A]^{2}$ in the exponent  
of  (\ref{E2}) by $\sum_{r}(4\lambda)^{-1}[w^{r}(t')-2\lambda \hat A^{r}]^{2}$.  The interaction picture state vector  in this case is ({\ref{E2}}) multiplied by $\exp( i \hat Ht)$:
\begin{equation}\label{E5}
|\psi,t\rangle_{w}={\cal T}e^{-(4\lambda)^{-1}\int_{0}^{t}dt'\sum_{r}[w^{r}(t')-2\lambda \hat A^{r}(t')]^{2}}|\psi,0\rangle, 
\end{equation}
\noindent where $ \hat A^{r}(t')\equiv\exp (i\hat Ht')\hat A^{r}\exp( -i\hat Ht')$.
The density matrix follows from (\ref{E5}), (\ref{E3}):
\begin{equation}\label{E6}
\hat \rho(t)\equiv\int P_{t}(w)Dw|\psi,t\rangle_{w} \thinspace_{w}\langle\psi,t|/P_{t}(w)=
{\cal T}e^{-\lambda/2\int_{0}^{t}dt'\sum_{r}[\hat A_{L}^{r}(t')-\hat A_{R}^{r}(t')]^{2}}\hat\rho(0)
\end{equation}
\noindent where $\hat A_{L}^{r}(t')$ ($\hat A_{R}^{r}(t')$)  appears to the left (right) of $\hat\rho(0)$, and is time-ordered (time reverse-ordered).  In the example described by (\ref{E4}), the density matrix (\ref{E6}) is
\[
\hat \rho(t)=\sum_{n,m}e^{-(\lambda t/2)(a_{n}-a_{m})^{2}}\alpha_{n}\alpha_{m}^{*}|a_{n}\rangle\langle a_{m}|,
\]
\noindent which encapsulates the ensemble's collapse behavior.

\subsection{CSL}\label{S3.3}

	The CSL proposal (Pearle 1989) is that collapse is engendered by distinctions between states at each point of space, so the index $r$ of $\hat A^{r}$ in (\ref{E5}) becomes  ${\bf x}$, 
\begin{equation}\label{E7}
|\psi,t\rangle_{w}={\cal T}e^{-(4\lambda)^{-1}\int_{0}^{t}\int dt'd{\bf x}'[w({\bf x}',t')-
2\lambda\hat  A({\bf x}',t')]^{2}}|\psi,0\rangle, 
\end{equation}
\noindent and the distinction looked at is mass density.  However, one cannot make the choice $\hat A({\bf x}, 0)=\hat M({\bf x})$, where $\hat M({\bf x})=\sum_{i}m_{i}\hat \xi_{i}^{\dagger}({\bf x})\hat \xi_{i}({\bf x})$ is the mass density operator ($m_{i}$ is the mass of the $i$th type of particle, so $m_{e}$, $m_{p}$, $m_{n}$... are the masses of  electrons, protons, neutrons...,  and $\hat \xi_{i}^{\dagger}({\bf x})$ is the creation operator for such a particle at location ${\bf x}$),   because this entails an infinite rate of energy increase of  particles ((\ref{E13}) with $a=0$).   Instead, adapting a ``gaussian smearing"  idea from the Ghirardi \textit{et al.} (1986) spontaneous localization (SL) model (see section 3.6), choose $\hat A^{{\bf x}}$ as, essentially,  proportional to the mass in a sphere of radius $a$ about ${\bf x}$:
\begin{equation}\label{E8}	
\hat A({\bf x,t})\equiv e^{i\hat Ht}\frac{1}{(\pi a^{2})^{3/4}}\int d{\bf z}\frac{\hat M({\bf z})}{m_{p}}e^{-(2a^{2})^{-1}({\bf x}-{\bf z})^{2}}e^{-i\hat Ht}
\end{equation}
\noindent The parameter value choices of SL, $\lambda \approx 10^{-16}$sec$^{-1}$ (according to (17),  the collapse rate for protons) and $a\approx 10^{-5}$cm are, so far, consistent with experiment (see section \ref{S3.4}), and will be adopted here.     
	  
	The density matrix associated with (\ref{E7}) is, as in (\ref{E6}),
\begin{equation}\label{E9}
\hat \rho(t)={\cal T}e^{-(\lambda/2)\int_{0}^{t}dt' d{\bf x}' [\hat A_{L}({\bf x}',t')-\hat A_{R}({\bf x}',t')]^{2}}\hat\rho(0), 
\end{equation}
\noindent which satisfies the differential equation 
\begin{equation}\label{E10}
\frac{d\hat \rho(t)}{dt}=-\frac{\lambda}{2}\int d{\bf x}' [\hat A({\bf x}',t),[\hat A({\bf x}',t),\hat\rho(t)]]
\end{equation}
\noindent of Lindblad-Kossakowski form.   

\subsection{Consequences of CSL}\label{S3.4}

	Since the state vector dynamics of CSL is different from that of standard quantum theory, there are phenomena for which the two make different predictions, allowing for experimental tests.  Consider an $N$-particle system with position operators $\hat X_{i}$
($\hat X_{i}|{\bf x}\rangle= x_{i}|{\bf x}\rangle$).  Substitution of $\hat A({\bf x}')$ from (\ref{E8}) in the Schr\"odinger picture version of (\ref{E10}), integration over ${\bf x}'$, and utilization of 
\[
f({\bf z})\hat M({\bf z})|{\bf x}\rangle=\sum_{i=1}^{N} m_{i} f(\hat{\bf X}_{i})\delta ({\bf z}-\hat{\bf X}_{i})|{\bf x}\rangle
\]
\noindent results in 
\begin{eqnarray}\label{E11}
\frac{d\hat\rho(t)}{dt}&=&-i[\hat\rho(t), \hat H]-
\frac{\lambda}{2}\sum_{i=1}^{N}\sum_{j=1}^{N}\frac{m_{i}}{m_{p}}\frac{m_{j}}{m_{p}}\nonumber\\
&&\times \bigg[ e^{-(4a^{2})^{-1}(\hat{\bf X}_{Li}-\hat{\bf X}_{Lj})^{2}}+
e^{-(4a^{2})^{-1}(\hat{\bf X}_{Ri}-\hat{\bf X}_ {Rj})^{2}} -2e^{-(4a^{2})^{-1}(\hat{\bf X}_{Li}-\hat{\bf 
X}_{Rj})^{2}}\bigg]\hat\rho(t)
\end{eqnarray}
\noindent which is a useful form for calculations first suggested by Pearle and Squires in 1994.
 
\subsubsection{Interference}\label{S3.4.1}
	Consider the collapse rate of  an initial state $|\phi\rangle=\alpha_{1}| 1\rangle+\alpha_{2}| 2\rangle$, where  $| 1\rangle$, $| 2 \rangle$ describe a clump of matter, of size $<<a$, at different locations with  separation  $>>a$.  Electrons may be neglected because of their small collapse rate compared to the much more massive nucleons, and the nucleon mass difference may be neglected.  In using (\ref{E11}) to calculate $d\langle 1|\hat\rho(t)|2\rangle/dt$, since 
$\exp[-(4a^{2})^{-1}(\hat{\bf X}_{i}-\hat{\bf X}_j)^{2}]\approx 1$ when acting on state $| 1\rangle$ or $| 2\rangle$, and $\approx 0$ when $\hat{\bf X}_{i}$ acts on $| 1\rangle$ and $\hat{\bf X}_j$ acts on $| 2\rangle$,  
(\ref{E11}) yields, for $N$ nucleons, the collapse rate $\lambda N^{2}$:
\begin{equation}\label{E12}
\frac{d\langle 1|\hat\rho(t)|2\rangle}{dt}=-i\langle 1|[\hat\rho(t), \hat H]|2\rangle-\lambda N^{2}\langle 1|\hat\rho(t)|2\rangle.
\end{equation}
\noindent If the clump undergoes a two-slit interference experiment, where the size and separation conditions above are satisfied for $\Delta T$sec, and if the result agrees with the standard quantum theory  prediction to 1\%, 
it also agrees with CSL provided $\lambda^{-1}>100N^{2}\Delta T$.  So far, interference  experiments with $N$ as large as $\approx 10^{3}$ have been performed  by Nairz, Arndt and Zeilinger in 2000.  
The SL value of $\lambda$ would be testable, that is, the quantum predicted interference pattern would be ``washed out" to 1\% accuracy,  if the clump were an $\approx 10^{-6}$cm radius sphere of mercury, which contains $N\approx 10^{8}$ nucleons, interfered for $ \Delta T=.01$sec. Currently envisioned but not yet performed experiments (e.g., by Marshall, Simon, Penrose and Bouwmester  in 2003) have been analyzed (e.g., by Bassi, Ippoliti and Adler in 2004 and by Adler in 2005) involving a superposition of a larger clump of matter in slightly displaced positions, entangled with a photon whose interference pattern is measured:  these proposed experiments are still too crude to detect  the SL value of $\lambda$, or 
 the gravitationally-based 
collapse rate proposed by Penrose in 1996 (see section \ref{Sec4} and papers by Christian in 1999 and  2005).

\subsubsection{Bound State Excitation}
	
	Collapse narrows wave packets, thereby imparting energy to particles.  If $\hat H=\sum_{i=1}^{N}\hat{\bf P}_{i}^{2}/2m_{i}+\hat V({\bf x}_{1},...{\bf x}_{N})$, it is straightforward to calculate from (\ref{E11}) that
\begin{equation}\label{E13}
\frac{d}{dt}\langle\hat H\rangle\equiv \frac{d}{dt}\hbox{Tr}[\hat H\hat\rho (t)]=\sum_{i=1}^{N}\frac{3\lambda \hbar^{2}}{4m_{i}a^{2}}.
\end{equation}
\noindent   For a nucleon, the mean rate of energy increase is quite small, $\approx 3\times10^{-25}$eV/sec. However, deviations from the mean can be significantly greater.  

	For, (\ref{E11}) predicts excitation of atoms and nucleii.  Let $|E_{0}\rangle$ be an initial bound energy eigenstate.  Expanding (\ref{E11}) in a power series in (bound state size/$a$)$^2{}$, the excitation rate of state $|E\rangle$ is 
\begin{equation}\label{E14}
\Gamma\equiv\frac{d\langle E|\hat\rho(t)|E\rangle}{dt}|_{t=0}=\frac{\lambda}{2a^{2}}\langle E|\sum_{i=1}^{N}\frac{m_{i}}{m_{p}}\hat{\bf X}_{i}|E_{0}\rangle{\bf \cdot}\langle E_{0}|\sum_{i=1}^{N}\frac{m_{i}}{m_{p}}\hat{\bf X}_{i}|E\rangle+\hbox{O(size}/a)^{4}.
\end{equation}
\noindent Since $|E_{0}\rangle$, $|E\rangle$ are eigenstates of the center of mass operator 
$\sum_{i=1}^{N}m_{i}\hat{\bf X}_{i}/\sum_{i=1}^{N}m_{i}$ with eigenvalue 0, the dipole contribution explicitly given in (\ref{E14}) vanishes identically.  This leaves the quadrupole contribution as the leading term, which is too small to be measured at present.

	However, the choice of $\hat A({\bf x})$ as mass density operator was made  only after experimental indication.  Let $g_{i}$ replace $m_{i}/m_{p}$ in (\ref{E11}), (\ref{E14}), so that $\lambda g^{2}_{i}$ is the 
collapse rate for the $i$th particle.  Then, experiments looking for the radiation expected from ``spontaneously" excited atoms and nucleii, in large amounts of matter for a long time, as shown by 
Collett, Pearle, Avignone and Nussinov in 1995, Pearle, Ring, Collar and Avignone in 1999 and Jones, Pearle and Ring in 2004, have placed the following limits: 

\[
 |g_{e}/g_{p}-m_{e}/m_{p}|<12m_{e}/m_{p}, \qquad |g_{n}/g_{p}-m_{n}/m_{p}|<3(m_{n}-m_{p})/m_{p}.   
\]

\subsubsection{Random Walk}
	According to (\ref{E7}), (\ref{E3}), the center of mass wave packet, of a piece of matter of size $\approx a$ or smaller, containing $N$ nucleons, achieves equilibrium size $s$ in a characteristic time $\tau_{s}$, and undergoes a random walk through a root  mean square distance $\Delta Q$:
	
\begin{equation}\label{E15}
s\approx \bigg[\frac{a^{2}\hbar}{\lambda m_{p}N^{3}}\bigg]^{1/4},\quad \tau_{s}\approx \frac{Nm_{p}s^{2}}{\hbar}, \quad \Delta Q\approx\frac{\hbar\lambda^{1/2}t^{3/2}}{m_{p}a}.  
\end{equation}
\noindent The results in  (\ref{E15}) were obtained by Collett and Pearle in 2003.  These quantitative results can be qualitatively understood as follows.

In time $\Delta t$, the usual Schr\"odinger equation expands a wave packet of size $s$ to 
$\approx s+(\hbar/Nm_{p}s)\Delta t$.  CSL collapse, by itself, narrows the wave packet to $\approx s[1-\lambda N^{2}(s/a)^{2}\Delta t]$.  The condition of no change in $s$ is the result quoted above.  $\tau_{s}$ is the time it takes the Schr\"odinger evolution to expand a wavepacket near size $s$ to  size $s$: $(\hbar/Nm_{p}s)\tau_{s}\approx s$. The $t^{3/2}$ dependence of  $\Delta Q$ arises because this is a random walk without damping (unlike Brownian motion, where $\Delta Q\sim t^{1/2}$).  The mean energy increase $\approx \lambda N\hbar^{2}m_{p}^{-1}a^{-2}t$ of (\ref{E13}) implies the root-mean-square velocity increase $\approx [\lambda \hbar^{2}m_{p}^{-2}a^{-2}t]^{1/2}$, whose product with $t$ is $\Delta Q$. 

For example, a sphere of density 1gm/cc and radius $10^{-5}$cm  has $s\approx 4\times10^{-7}$cm, 
$\tau_{s}\approx 0.6$sec and $\Delta Q\approx 5[t \hbox { in days}]^{3/2}$cm. At the reported achieved low pressure of $5\times 10^{-17}$Torr at $4.2^{\circ}$K  reported by Gabrielse's group in 1990, 
the mean collision time with gas molecules is $\approx 80$min, over which  $\Delta Q\approx  0.7$mm.    
Thus, observation of this effect should be feasible.

\subsection{Further Remarks}
	It is possible to define energy for the $w({\bf x},t)$ field so that total  energy is conserved: as the particles gain energy, the $w$-field loses energy, as shown by Pearle in 2005.
	   
	Attempts to construct a special relativistic CSL-type model have not yet succeeded although Pearle in 1990, 1992, 1999,  Ghirardi, Grassi and Pearle in 1990 and  Nicrosini and Rimini in 2003 have made valiant attempts.  The problem is that the white noise field $w({\bf x},t)$ contains all wavelengths and frequencies, exciting the vacuum in lowest order in $\lambda$ to produce particles at the unacceptable rate of infinite energy/sec-cc.  Collapse models which utilize a colored noise field $w$ have a similar problem  
in higher order. In 2005, Pearle suggested a ``quasi-relativistic" model which reduces to CSL in the low speed limit.

	CSL is a phenomenological model which describes dynamical collapse so as to achieve S.  Besides needing decisive experimental verification, it needs identification of the $w({\bf x},t)$ field with a physical entity. 
	
	Other collapse models which have been investigated are briefly described below.  
	
\subsection{Spontaneous Localization Model (SL)}\label{S3.6}

	The SL model of Ghirardi \textit{et al.} (1986), although superseded by CSL, is historically important and conceptually valuable.  Let $\hat H=0$ for simplicity, and consider a single particle whose wave function at time $t$ is $\psi({\bf x},t)$.  Over the next interval $dt$, with probability $1-\lambda dt$, it does not change.  With probability $\lambda dt$ it does change, by being ``spontaneously localized" or ``hit."  A hit means that the 
new (unnormalized) wavefunction suddenly becomes
\[
\psi({\bf x},t+dt)=\psi({\bf x},t)(\pi a^{2})^{-3/4}e^{-(2a^{2})^{-1}({\bf x}-{\bf z})^{2}}\quad \hbox{with probability}\quad \lambda dtd{\bf z}\int d{\bf x}|\psi({\bf x},t+dt)|^{2}.
\]
\noindent Thus ${\bf z}$, the ``center" of the hit, is most likely to be located where the 
wavefunction is large.  For a single particle in the superposition described in section \ref{S3.4.1}, 
a single hit is overwhelmingly likely to reduce the wave function to one or the other location, with total 
probability $|\alpha_{i}|^{2}$, at the rate $\lambda$.  

For an $N$ particle clump, it is considered that each particle has the same independent probability, $\lambda dt$, of being hit.  But,  for the example in section \ref{S3.4.1}, a single hit on any particle in one location of the clump has the effect of multiplying the wave function part describing the clump in the other location by the tail of the gaussian, thereby collapsing the wave function at the rate 	$\lambda N$.

	By use of the gaussian hit rather than a delta function hit, SL solves the problem of giving too much energy to particles as mentioned in section \ref{S3.3}.  It also solves the problem of achieving a slow collapse rate for a superposition of small objects and a fast collapse rate for a superposition of large objects.  However, the SL hits on individual particles destroys the (anti-) symmetry of wave functions.  The CSL collapse toward mass density eigenstates removes that problem.   Also,  while SL modifies the 
Schr\"odinger evolution of a wave function, it involves discontinuous dynamics and so is not described by a modified Schr\"odinger equation as is CSL.  
		
\section{Other Models}\label{Sec4}	

For a single (low-energy) particle, the polar decomposition $\Psi
=Re^{(i/\hbar )S}$ of the Schr\"{o}dinger equation implies two real
equations,%
\begin{equation}
\frac{\partial R^{2}}{\partial t}+\mathbf{\nabla }\cdot (R^{2}\frac{\mathbf{%
\nabla }S}{m})=0  \label{S1}
\end{equation}%
(the continuity equation for $R^{2}=\left\vert \Psi \right\vert ^{2}$) and%
\begin{equation}
\frac{\partial S}{\partial t}+\frac{(\mathbf{\nabla }S)^{2}}{2m}+V+Q=0
\label{S2}
\end{equation}%
where $Q\equiv -(\hslash ^{2}/2m)\nabla ^{2}R/R$ is the \textquotedblleft
quantum potential\textquotedblright . (These equations have an obvious
generalisation to higher-dimensional configuration space.) In 1926, Madelung
proposed that one should start from (\ref{S1}) and (\ref{S2}) -- regarded as
hydrodynamical equations for a classical charged fluid with mass density $%
mR^{2}$ and fluid velocity $\mathbf{\nabla }S/m$ -- and construct $\Psi
=Re^{(i/\hbar )S}$ from the solutions.

This \textquotedblleft hydrodynamical\textquotedblright\ interpretation
suffers from many difficulties, especially for many-body systems. In any
case, a criticism by Wallstrom (1994) seems decisive: (\ref{S1}) and (\ref%
{S2}) (and their higher-dimensional analogs) are not, in fact, equivalent to
the Schr\"{o}dinger equation. For, as usually understood, the quantum wave
function $\Psi $ is a single-valued and continuous complex field, which
typically possesses nodes ($\Psi =0$), in the neighborhood of which the
phase $S$ is multivalued, with values differing by integral multiples of $%
2\pi \hslash $. If one allows $S$ in (\ref{S1}), (\ref{S2}) to be
multi-valued, there is no reason why the allowed values should differ by
integral multiples of $2\pi \hslash $, and in general $\Psi $ will not be
single-valued. On the other hand, if one restricts $S$ in (\ref{S1}), (\ref%
{S2}) to be single-valued, one will exclude wave functions -- such as those
of nonzero angular momentum -- with a multivalued phase. (This problem does
not exist in pilot-wave theory as we have presented it here, where $\Psi $
is regarded as a basic entity.)

Stochastic mechanics, introduced by F\'{e}nyes in 1952 and Nelson (1966),
has particle trajectories $\mathbf{x}(t)$ obeying a \textquotedblleft
forward\textquotedblright\ stochastic differential equation $d\mathbf{x}(t)=%
\mathbf{b}(\mathbf{x}(t),t)dt+d\mathbf{w}(t)$, where $\mathbf{b}$ is a drift
(equal to the mean forward velocity) and $\mathbf{w}$ a Wiener process, and
also a similar \textquotedblleft backward\textquotedblright\ equation.
Defining the \textquotedblleft current velocity\textquotedblright\ $\mathbf{v%
}=\frac{1}{2}(\mathbf{b}+\mathbf{b}_{\ast })$, where $\mathbf{b}_{\ast }$ is
the mean backward velocity, and using an appropriate time-symmetric
definition of mean acceleration, one may impose a stochastic version of
Newton's second law. If one assumes, in addition, that $\mathbf{v}$ is a
gradient ($\mathbf{v}=\mathbf{\nabla }S/m$ for some $S$), then one obtains (%
\ref{S1}), (\ref{S2}) with $R\equiv \sqrt{\rho }$, where $\rho $ is the
particle density. Defining $\Psi \equiv \sqrt{\rho }e^{(i/\hbar )S}$, it
appears that one recovers the Schr\"{o}dinger equation for the derived
quantity $\Psi $. However, again, there is no reason why $S$ should have the
specific multivalued structure required for the phase of a single-valued
complex field. It then seems that, despite appearances, quantum theory
cannot in fact be recovered from stochastic mechanics (Wallstrom 1994). The
same problem occurs in models that use stochastic mechanics as an
intermediate step (e.g., Markopoulou and Smolin in 2004): the Schr\"{o}%
dinger equation is obtained only for exceptional, nodeless wave functions.

	Bohm and Bub (1966) first proposed dynamical wave function collapse through deterministic evolution.  Their collapse outcome is determined by the value of a Wiener-Siegel hidden variable (a variable distributed uniformly over the unit hypersphere in a Hilbert space identical to that of the statevector).  
In 1976, Pearle  proposed dynamical wave function collapse equations where the collapse outcome is determined by a random variable, and suggested (Pearle 1979) that the modified Schr\"odinger equation be formulated as an It\^o stochastic differential equation, a suggestion which has been widely followed. (The equation for the state vector given here, which is physically more transparent, has its time derivative equivalent to a Stratonovich stochastic differential equation, which is readily converted to the It\^o form.)  The importance of having the density matrix describing collapse be of the Lindblad-Kossakowski form was emphasized by Gisin in 1984 and Diosi in 1988.  The stochastic differential  Schr\"odinger equation which achieves this was found independently by Diosi in 1988 and by Belavkin, Gisin and Pearle  in separate papers in 1989 (see also Ghirardi \textit{et al.} 1990).  

	A gravitationally motivated stochastic collapse dynamics was proposed by Diosi in 1989 (and somewhat corrected by Ghirardi \textit{et al.} in 1990).  Penrose  emphasized in 1996 that a quantum state, such as that describing a mass in a superposition of two places, puts the associated space-time geometry also in a superposition, and has argued that this should lead to wave function collapse. He suggests that the collapse time should be $\sim \hbar/\Delta E$, where $\Delta E$ is the gravitational potential energy change obtained by actually displacing two such masses:  for example, the  collapse time $\approx \hbar /(Gm^{2}/R)$, where the mass is $m$,  its size is $R$, and the displacement is $\approx R$ or larger.  No specific dynamics is offered, just the vision that this will be a property of a correct future quantum theory of gravity. 

	Collapse to energy eigenstates was first proposed by Bedford and Wang in 1975 and 1977 and, in the context of stochastic collapse (e.g.,  (11) with $\hat A=\hat H$), by Milburn in 1991 and Hughston in 1996, but it has been argued by Finkelstein in 1993 and Pearle in 2004 that such energy-driven collapse cannot give a satisfactory picture of the macroscopic world.  Percival in 1995 and in a 1998 book,  and Fivel in 1997 have discussed energy-driven collapse for microscopic situations.  
	
	Adler (2004) has presented a classical theory (a hidden-variables theory)  from which it is argued that quantum theory  ``emerges"  at the ensemble level. The classical variables are $N\times N$ matrix field amplitudes at points of space.  They obey appropriate classical Hamiltonian dynamical equations which he calls ``trace dynamics," since the Hamiltonian, Lagrangian, Poisson bracket, etc. expressions have the form of the trace of products of matrices and their sums with constant coefficients.  Using  classical  statistical mechanics, canonical ensemble averages of (suitably projected) products of fields are analyzed  and it is argued that they obey all the properties associated with Wightman functions, from which quantum field theory, and its non-relativistic limit quantum mechanics, may be derived.  As well as obtaining the algebra of quantum theory in this way, it is argued that statistical fluctuations around the canonical ensemble can give rise to wave function collapse behavior, of the kind discussed here, both energy-driven and CSL-type mass-density-driven collapse.  The Hamiltonian needed for this theory to work is not provided but, as the argument progresses, its necessary features are delimited.

\begin{center}
\textbf{BIBLIOGRAPHY}
\end{center}

Adler SL (2004) \textit{Quantum Theory as an Emergent Phenomenon}. Cambridge
University Press, Cambridge.

Bassi A and Ghirardi GC (2003) Dynamical reduction models. \textit{Physics
Reports} 379: 257--426. ArXiv: quant-ph/0302164.

Bell JS (1987) \textit{Speakable and Unspeakable in Quantum Mechanics}.
Cambridge University Press, Cambridge.

Bohm D (1952) A suggested interpretation of the quantum theory in terms of
`hidden' variables. I and II. \textit{Physical Review} 85: 166--179; 180--193.

Bohm D and Bub J (1966) A proposed solution of the measurement problem 
in quantum mechanics by a hidden variable theory. \textit{Reviews of Modern Physics} 38: 453--469. 

D\"{u}rr D, Goldstein S and Zangh\`{\i} N (2003) Quantum equilibrium and the
role of operators as observables in quantum theory. ArXiv: quant-ph/0308038.

Ghirardi G, Rimini A and Weber T (1986) Unified dynamics for microscopic and macroscopic systems.
 \textit{Physical Review D} 34: 470--491. 
 
 Ghirardi G, Pearle P and Rimini A (1990) Markov processes in Hilbert space and continuous 
 spontaneous localization of systems of identical particles. \textit{Physical Review A} 42: 78--89.
 
Holland PR (1993) \textit{The Quantum Theory of Motion: an Account of the de
Broglie-Bohm Causal Interpretation of Quantum Mechanics}. Cambridge University
Press, Cambridge.

Nelson E (1966) Derivation of the Schr\"{o}dinger equation from Newtonian
mechanics. \textit{Physical Review} 150: 1079--1085.

Pearle P (1979) Toward explaining why events occur.  \textit{International Journal of Theoretical Physics} 18: 489--518. 

Pearle P (1989)  Combining stochastic dynamical state-vector reduction with spontaneous localization \textit{Physical Review A} 39: 2277--2289. 

Pearle P (1999) Collapse models. In: Petruccione F and Breuer HP (eds.)
\textit{Open Systems and Measurement in Relativistic Quantum Theory}, pp
195--234. Springer-Verlag, Heidelberg.  ArXiv: quant-ph/9901077.

Valentini A (1991) Signal-locality, uncertainty, and the subquantum
\textit{H}-theorem. I and II. \textit{Physics Letters} A156: 5--11; A158: 1--8.

Valentini A (2002a) Signal-locality in hidden-variables theories.
\textit{Physics Letters} A297: 273--278.

Valentini A (2002b) Subquantum information and computation. \textit{Pramana --
Journal of Physics} 59: 269--277.  ArXiv: quant-ph/0203049.

Valentini A (2004a) Black holes, information loss, and hidden variables.
ArXiv: hep-th/0407032.

Valentini A (2004b) Universal signature of non-quantum systems.
\textit{Physics Letters} A332: 187--193.  ArXiv: quant-ph/0309107.

Valentini A and Westman H (2005) Dynamical origin of quantum probabilities.
\textit{Proceedings of the Royal Society of London} A461: 253--272.

Wallstrom TC (1994) Inequivalence between the Schr\"{o}dinger equation and the
Madelung hydrodynamic equations. \textit{Physical Review} A49: 1613--1617.

\begin{center}
-----------------
\end{center}

\end{document}